\documentclass[sn-nature]{sn-jnl}

\usepackage{graphicx}%
\usepackage{multirow}%
\usepackage{amsmath,amssymb,amsfonts}%
\usepackage{amsthm}%
\usepackage{mathrsfs}%
\usepackage[title]{appendix}%
\usepackage{xcolor}%
\usepackage{textcomp}%
\usepackage{manyfoot}%
\usepackage{booktabs}%
\usepackage{algorithm}%
\usepackage{algorithmicx}%
\usepackage{algpseudocode}%
\usepackage{listings}%
\usepackage{hyperref}

\usepackage{parskip}
\usepackage[capitalise,noabbrev,nameinlink]{cleveref}

\usepackage{lineno}
\usepackage{xr}
\usepackage{xr-hyper}
\usepackage{hyperref}

\begin{document}

\title[Article Title]{Evidential Deep Learning for Interatomic Potentials}


\author[1,3]{\fnm{Han} \sur{Xu}}
\equalcont{These authors contributed equally to this work.}
\author[1,4]{\fnm{Taoyong} \sur{Cui}}
\equalcont{These authors contributed equally to this work.}
\author[1]{\fnm{Chenyu} \sur{Tang}}
\equalcont{These authors contributed equally to this work.}
\author[1,7]{\fnm{Jinzhe} \sur{Ma}}
\author[1]{\fnm{Dongzhan} \sur{Zhou}}
\author[1]{\fnm{Yuqiang} \sur{Li}}
\author[3]{\fnm{Xiang} \sur{Gao}}
\author[5,6]{\fnm{Xingao} \sur{Gong}}
\author[1]{\fnm{Wanli} \sur{Ouyang}}
\author*[1]{\fnm{Shufei} \sur{Zhang}}\email{zhangshufei@pjlab.org.cn}
\author*[1,2]{\fnm{Mao} \sur{Su}}\email{sumao@pjlab.org.cn}

\affil[1]{\orgname{Shanghai Artificial Intelligence Laboratory}, \orgaddress{\city{Shanghai}, \postcode{200232}, \country{China}}}

\affil[2]{\orgname{Shenzhen Institute of Advanced Technology, Chinese Academy of Sciences}, \orgaddress{\city{Shenzhen}, \postcode{518055}, \country{China}}}

\affil[3]{\orgdiv{The State Key Laboratory of Chemical Engineering, College of Chemical and Biological Engineering}, \orgname{Zhejiang University}, \orgaddress{\city{Hangzhou}, \postcode{310027}, \country{China}}}

\affil[4]{\orgname{The Chinese University of Hong Kong}, \orgaddress{\city{Hong Kong}, \postcode{999077}, \country{China}}}

\affil[5]{\orgdiv{Key Laboratory for Computational Physical Sciences (MOE), State Key Laboratory of Surface Physics, Department of Physics}, \orgname{Fudan University}, \orgaddress{\city{Shanghai}, \postcode{200433}, \country{China}}}

\affil[6]{\orgname{Shanghai Qi Zhi Institute}, \orgaddress{\city{Shanghai}, \postcode{200232}, \country{China}}}

\affil[7]{\orgdiv{School of Physical Science and Technology}, \orgname{ShanghaiTech University}, \orgaddress{\city{Shanghai}, \postcode{201210}, \country{China}}}


\abstract{
Machine learning interatomic potentials (MLIPs) have been widely used to facilitate large-scale molecular simulations with accuracy comparable to ab initio methods. In practice, MLIP-based molecular simulations often encounter the issue of collapse due to reduced prediction accuracy for out-of-distribution (OOD) data. Addressing this issue requires enriching the training dataset through active learning, where uncertainty serves as a critical indicator for identifying and collecting OOD data. However, existing uncertainty quantification (UQ) methods tend to involve either expensive computations or compromise prediction accuracy. In this work, we introduce evidential deep learning for interatomic potentials (eIP) with a physics-inspired design. Our experiments indicate that eIP provides reliable UQ results without significant computational overhead or decreased prediction accuracy, consistently outperforming other UQ methods across a variety of datasets. Furthermore, we demonstrate the applications of eIP in exploring diverse atomic configurations, using examples including water and universal potentials. These results highlight the potential of eIP as a robust and efficient alternative for UQ in molecular simulations.
}


\maketitle

\section{Introduction}\label{sec1}

Molecular dynamics (MD) simulation provides atomic insights into physical and chemical processes and has become an indispensable research tool in computational physical science~\cite{mccammon1977dynamics, karplus2002molecular, warshel2002molecular}. Classical MD simulation uses an empirical potential function to determine interatomic forces~\cite{cornell1995second, mackerell1998all}, which is computationally efficient but not accurate enough, especially when polarization or many-body interactions are important~\cite{unke2021machine}. In contrast, ab initio approach for modeling atomic interactions is based solely on fundamental physical principles, leading to generally higher accuracy and transferability~\cite{car1985unified, huang2023central}, but the high computational cost limits the size of systems that can be simulated. To achieve both efficiency and accuracy, machine learning interatomic potentials (MLIPs) have been proposed~\cite{ butler2018machine, noe2020machine, manzhos2020neural, keith2021combining}, which allows to learn ab initio interatomic potentials and perform MD simulations with much lower computational cost. MLIPs have been successfully applied in the study of amorphous solid~\cite{deringer2021origins}, catalysis~\cite{galib2021reactive}, chemical reaction~\cite{zeng2020complex}, and more.

One of the primary challenges to MLIP-based MD simulations lies in the construction of the training dataset, which should include various configurations that may appear during the simulation. Inadequate training data will lead to decreased accuracy or even failure of the simulations~\cite{fu2022forces, cui2024online}. This challenge limits the application of MLIP-based MD simulations. Active learning based on uncertainty quantification (UQ) plays a crucial role in constructing training sets for MLIPs~\cite{smith2018less, zhang2020dp, yuan2023active, moon2024active}. During active learning, configurations with higher uncertainties are sampled to enrich the training set. This process usually needs to be repeated dozens or more times~\cite{zhang2020dp}, and the computational cost required for UQ could be considerable. Therefore, a robust yet efficient method for UQ is desired.

A variety of UQ methods have been developed for MLIPs. 
Moment tensor potential~\cite{novikov2020mlip} uses an extrapolation parameter to estimate uncertainty, but this method does not apply to deep neural network models. 
Gaussian approximation potential~\cite{bartok2015g} utilizes Gaussian process regression to provide UQ along with its predictions. However, the primary limitation of Gaussian approximation potential lies in its computational cost, which scales cubically with the dataset size. 
Ensemble methods~\cite{lakshminarayanan2017simple} are quite reliable for UQ, but also suffer from computational burdens due to the training of multiple models. 
Single-model methods, such as Monte Carlo dropout~\cite{gal2016dropout, wen2020uncertainty,thaler2024active}, Gaussian mixture models (GMM)~\cite{zhu2023fast}, and mean-variance estimation (MVE)~\cite{nix1994estimating}, mitigate the computational issue, but their performances are still not satisfactory~\cite{tan2023single}.

Evidential deep learning~\cite{amini2020deep,soleimany2021evidential} is a promising alternative, which estimates uncertainty through a single forward pass and requires minimal extra computational resources. Another advantage of evidential deep learning is that it can estimate aleatoric and epistemic uncertainties separately. 
Aleatoric uncertainty arises from intrinsic noise in the data and cannot be evaded or reduced. In contrast, epistemic uncertainty reflects the fidelity of the model in its representation of the data (excluding aleatoric effects) and decreases as the number of training samples increases~\cite{hullermeier2021aleatoric}.
The ability of evidential deep learning to distinguish between these two types of uncertainty is particularly beneficial for active learning, where we want to sample data with high epistemic uncertainty rather than aleatoric uncertainty. 
However, recent attempts~\cite{tan2023single,wollschlager2023uncertainty} trying to integrate evidential deep learning with MLIPs result in unsatisfactory performance. Failures may be attributed to inappropriate design in model architecture.

In this work, we reexamine the uncertainty associated with MLIPs from a physical perspective and propose a framework for UQ based on evidential deep learning. 
We call this framework the evidential interatomic potential (eIP). The performance of eIP is evaluated across various datasets and benchmarked with other UQ methods, demonstrating outstanding performance with minimal additional computational cost. 
Then, we extend the application of eIP to active learning and uncertainty-driven dynamics (UDD) simulations~\cite{kulichenko2023uncertainty}, enabling efficient exploration of the diverse atomic configurations. 
Lastly, we train a universal potential using eIP and achieve real-time UQ during simulations, which is challenging for ensemble-based methods due to their computational complexity.

\section{Results}\label{sec2}

\subsection{Preliminary}\label{}

\textbf{Machine learning interatomic potential (MLIP).}
MLIPs are used to predict energy and forces within a given atomic configuration. For a system comprising $N$ atoms, MLIPs typically take the atomic species $Z \in \mathbb{R}^{N}$ and coordinates $R \in \mathbb{R}^{N\times3}$ as input and outputs the total potential energy $E$. The forces $F \in \mathbb{R}^{N\times3}$ exerted on the atoms are derived by calculating the negative gradient of $E$ with respect to the coordinates. The primary distinction among various MLIPs lies in the algorithm used to convert the input information into vectorized features that represent the local atomic environments. These features are designed to be invariant or equivariant under translation, rotation, and permutation. 

\textbf{Aleatoric and epistemic uncertainty.} 
Two categories of uncertainty can be modeled in deep learning. Aleatoric uncertainty arises from noise in data labels, while epistemic uncertainty arises from inaccurate predictions due to data scarcity. In the study of MLIPs, noise in data labels can be eliminated through strict ab initio calculations, although inappropriate calculation settings may introduce noise. In practice, MLIPs often suffer from epistemic uncertainty, which can be mitigated by adding more training data through active learning. 
For the sake of simplicity, the term "uncertainty" mentioned in the following experimental results refers to epistemic uncertainty. We will discuss aleatoric uncertainty in Supplementary Section S2. 

\textbf{Evidential deep learning.} 
Evidential deep learning is an efficient method to estimate the uncertainty of the results predicted by neural networks. Starting from a maximum likelihood perspective, the targets are assumed to be drawn from a Gaussian distribution but with unknown mean and variance $(\mu,\sigma^2)$. A Gaussian prior is placed on the unknown mean $\mu$ and an Inverse-Gamma prior on the unknown variance $\sigma^2$, leading to the Normal Inverse-Gamma distribution with a set of parameters $\mathbf{m}=(\gamma, \nu, \alpha, \beta)$. Neural networks are then trained to infer $\mathbf{m}$, and the prediction, aleatoric, and epistemic uncertainty are calculated as~\cite{amini2020deep}:
\begin{equation}
\underbrace{\mathbb{E}[\mu] = \gamma}_\text{prediction}, \quad
\underbrace{\mathbb{E}[\sigma^2] = \frac{\beta}{\alpha-1}}_\text{aleatoric}, \quad
\underbrace{\text{Var}[\mu] = \frac{\beta}{\nu (\alpha-1)}}_\text{epistemic}.
\end{equation}

\subsection{Framework of eIP}\label{}

\begin{figure}[ht]
    \centering
    \includegraphics[width=13cm]{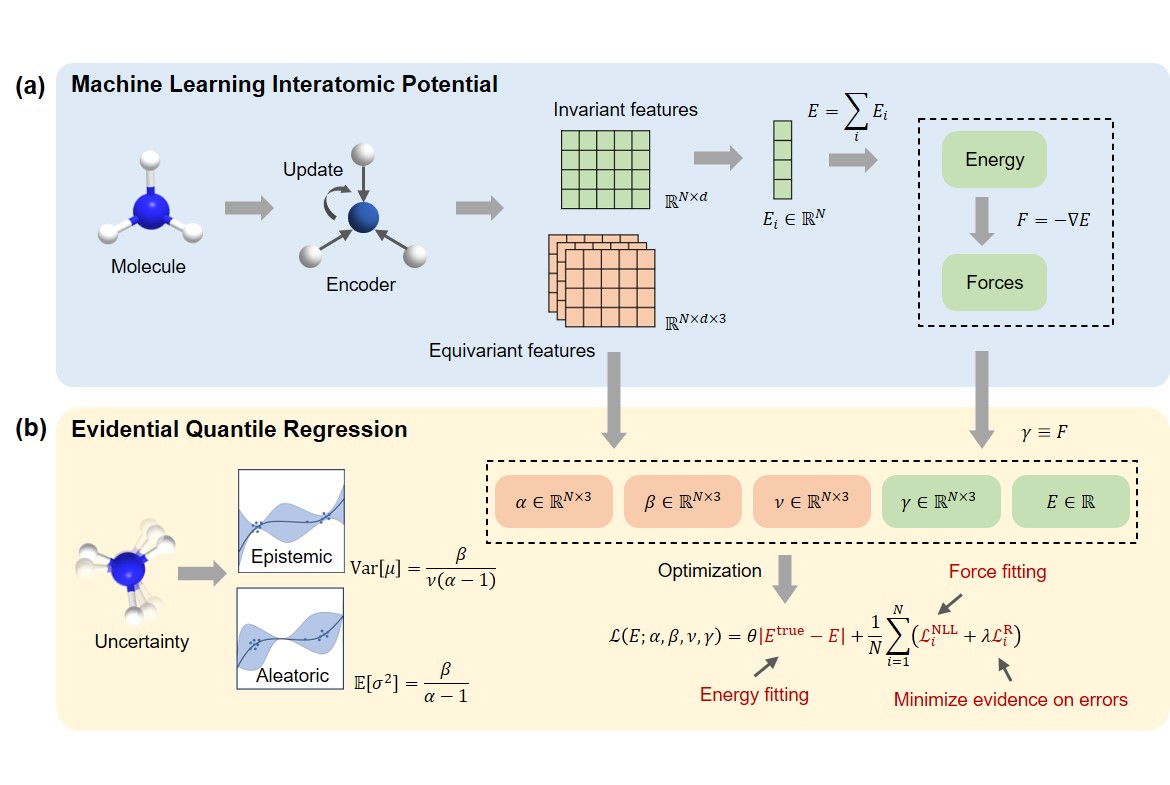}
    \caption{\textbf{Framework of eIP.} (a) A typical equivariant interatomic potential model extracting both invariant and equivariant features. The invariant features are used to output the potential energy. (b) Evidential quantile regression. The equivariant features are used to output the parameters for uncertainty quantification.}
    \label{fig:eip}
\end{figure}

The framework of eIP consists of an MLIP block for energy and force prediction, and an evidential quantile regression block for UQ, as illustrated in \cref{fig:eip}. In designing eIP, we have considered the following points, which are indispensable for achieving robust performances. 

\textbf{Locality.} In most MLIPs, the potential energy is calculated as the sum of atomic contributions, $E=\sum_{i=1}^N E_i$, with the model learning the mapping from the local environment of the atom $i$ to $E_i$. Therefore, we estimate the uncertainty associated with $E_i$ rather than the total potential energy $E$. However, we do not have the ground truth for $E_i$. Fortunately, we can adapt the atomic forces instead of $E_i$ to estimate the uncertainty per atom. 

\textbf{Directionality.} 
We attribute the occurrence of uncertainty in MLIP predictions to inadequate learning of local atomic configurations. Consequently, the uncertainty should be directionally dependent. This point is illustrated using a three-atom toy system in Supplementary Section S1. 
In the following experiments, we employ the equivariant backbone PaiNN~\cite{schutt2021equivariant} to extract equivariant features and output the parameters of the Normal Inverse-Gamma prior distribution, but eIP applies to other equivariant backbones. 

\textbf{Quantile regression.} Evidential deep learning assumes that the targets are drawn from a Gaussian distribution, which may not adequately describe the target distribution of MLIPs. To alleviate this limitation, we employ the Bayesian quantile regression model~\cite{huttel2023deep}, which improves upon the original evidential deep learning and yields better performance for non-Gaussian distributions. The calculation procedure of Bayesian quantile regression is similar to that of evidential deep learning, but the parameters $\mathbf{m}$ are optimized with different loss functions.

\subsection{Experiments}\label{}

\begin{figure}[ht]
    \centering
    \includegraphics[width=13cm]{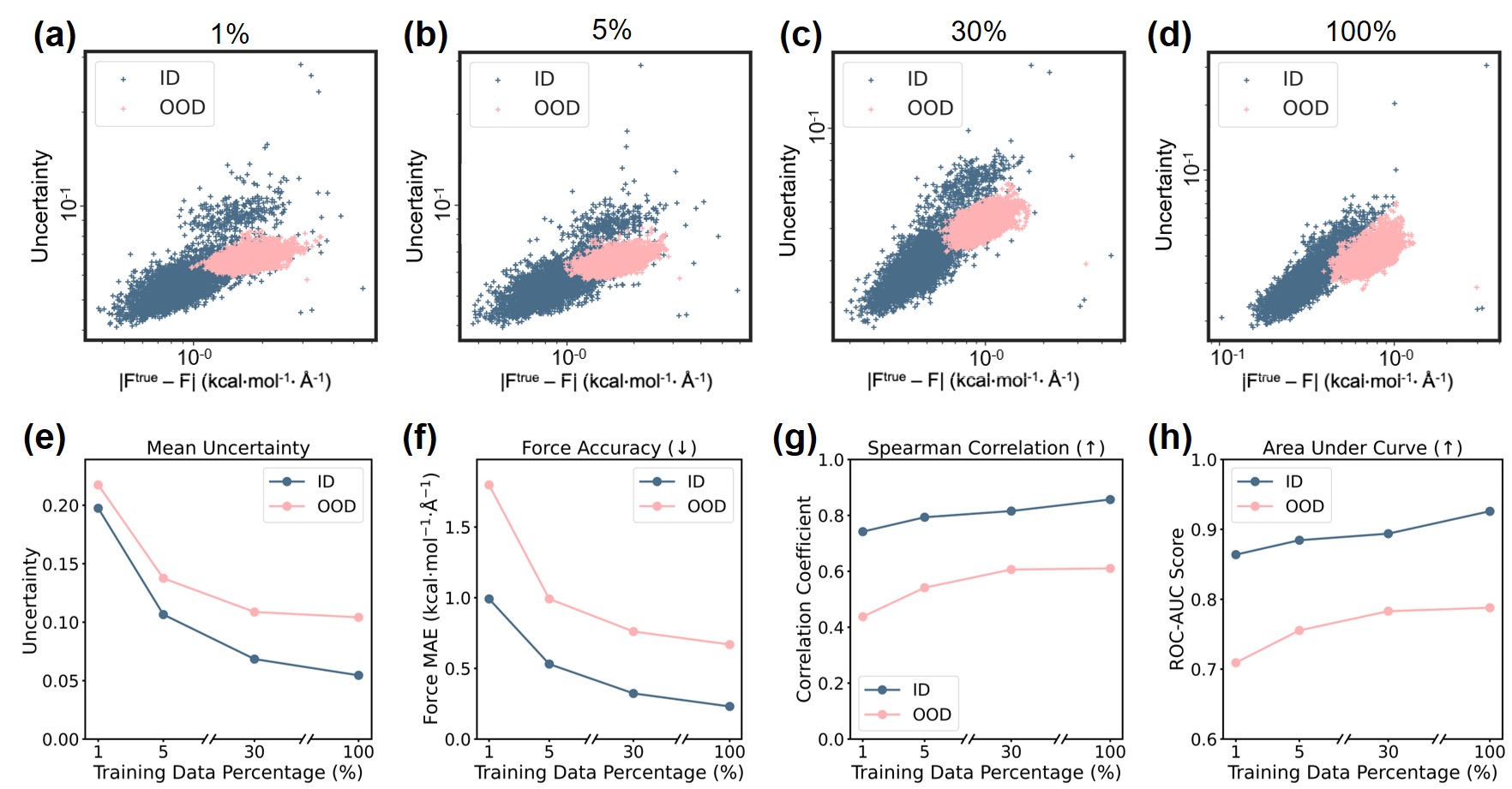}
    \caption{\textbf{Results on ISO17 dataset with increasing data volume.} (a)-(d) Scatter plots of uncertainties versus force errors using 1\%, 5\%, 30\%, and 100\% of the training data, respectively. Each point corresponds to the averaged uncertainty/error in a molecule. (e) Mean uncertainty on the test set. (f) Force mean absolute errors (MAEs) on the test set. (g) Spearman's rank correlation coefficients between uncertainty and force error. (h) ROC-AUC scores.}
    \label{fig:iso17}
\end{figure}

\textbf{ISO17 dataset.}
We started by assessing the performance of eIP using the ISO17 dataset, which comprises MD trajectories of $\text{C}_{7}\text{O}_{2}\text{H}_{10}$ isomers. This dataset is divided into in-distribution (ID) and out-of-distribution (OOD) subsets, making it particularly suitable for uncertainty quantification (UQ). 
In the ID scenario (known molecules/unknown conformations), the test molecules are also present in the training set. In contrast, the OOD scenario (unknown molecules/unknown conformations) involves test molecules that are not in the training set. The training set contains 400,000 conformations, which is a substantial amount for such small molecules. Therefore, we also explore the impact of training data volume. Specifically, we train the model using 1\%, 5\%, 30\%, and 100\% of the training data, respectively. 
\cref{fig:iso17}(a)-(d) show the scatter plots that compare uncertainties with force errors for different amounts of training data, demonstrating positive correlations in both ID and OOD scenarios. 
The mean uncertainty and mean absolute error (MAE) for force predictions are shown in \cref{fig:iso17}(e) and (f), respectively. As expected, both metrics decrease with an increase in the amount of training data. Furthermore, we evaluated the reliability of UQ using additional metrics, including Spearman's rank correlation coefficient and the area under the receiver operating characteristic curve (ROC-AUC). As shown in \cref{fig:iso17}(g) and (h), both Spearman's rank correlation coefficient and ROC-AUC improve as the amount of training data grows. In the ID scenario, Spearman's rank correlation coefficients ranging from 0.74 to 0.86 and ROC-AUC values ranging from 0.86 to 0.93 indicate the strong performance of eIP. In the OOD scenario, although the molecules in the test set are absent from the training set, the metrics remain within favorable ranges, highlighting the robustness of eIP.

\begin{figure}[ht]
    \centering
    \includegraphics[width=13cm]{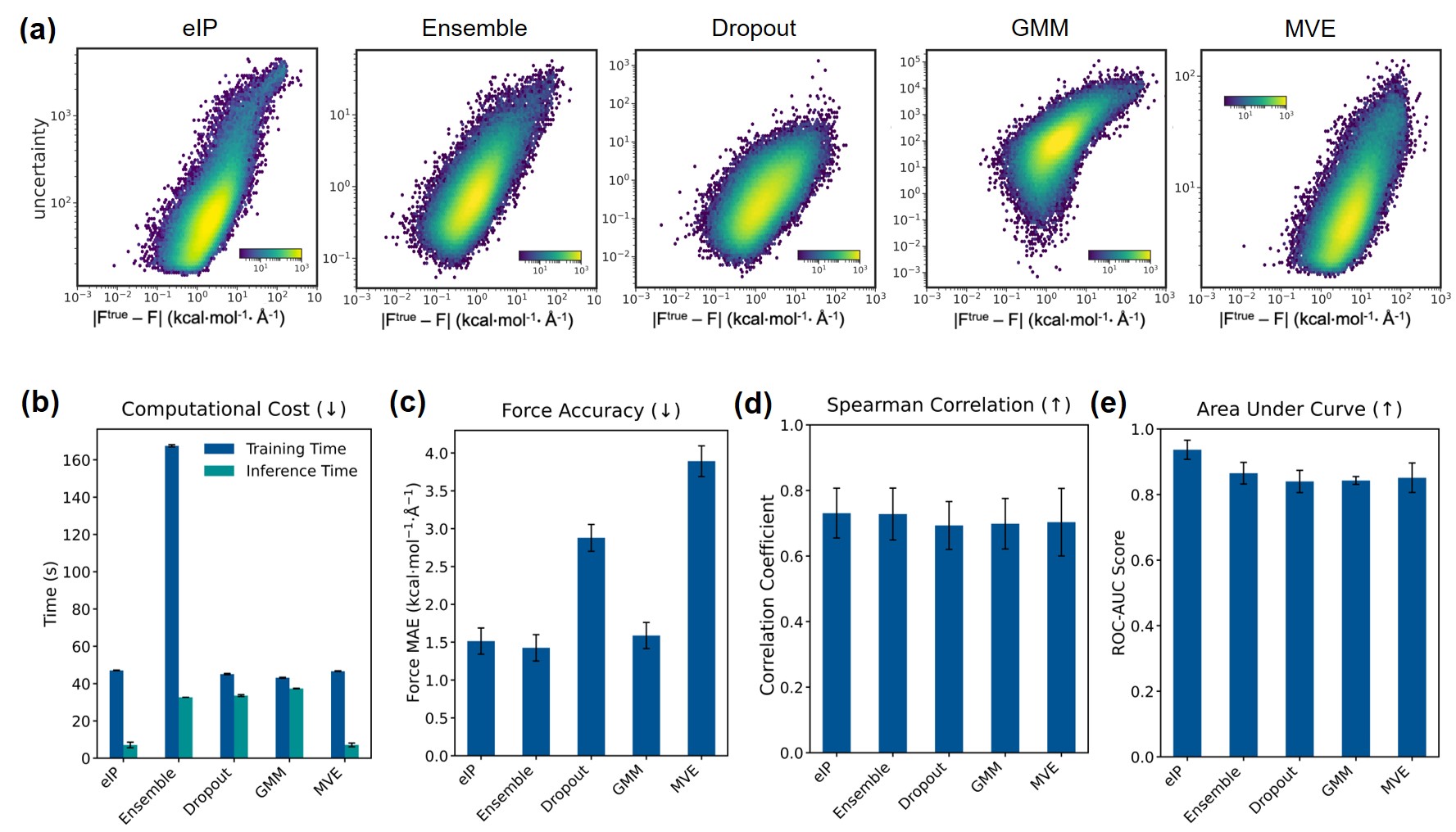}
    \caption{\textbf{Comparing eIP with other uncertainty quantification methods on silica glass dataset.} (a) Hexbin plots of uncertainties versus atomic force errors. (b) Computational costs. The "training time" here refers to the time required for each epoch. The "inference time" includes the time cost of computing uncertainty. (c) Force mean absolute errors (MAEs) on the test set. (d) Spearman's rank correlation coefficients between uncertainty and force error. (e) ROC-AUC scores.
    While all five methods achieve strong Spearman's rank correlations and ROC-AUC scores, ensemble, dropout, and GMM require longer computation times; dropout and MVE exhibit much lower accuracy in force prediction.
    }
    \label{fig:silica}
\end{figure}

\textbf{Silica glass dataset.}
We then evaluate eIP's performance for more complex systems using a silica glass dataset, which comprises large bulk structures. Given the challenges in partitioning large structures into ID and OOD datasets, we adopted the dataset partition scheme consistent with the previous study~\cite{tan2023single}. 
We also compare eIP with other UQ methods, including ensemble, Monte Carlo dropout, Gaussian mixture model (GMM), and Mean-variance estimation (MVE), whose implementations are provided in Supplementary S5. 
\cref{fig:silica}(a) shows the scatter plots of uncertainties versus force errors and indicates that all methods achieve positive correlations. 
\cref{fig:silica}(b) presents the computational efficiency analysis of the five methods. Despite the good performance of the ensemble method, it requires four times the training time of the other methods due to training four independent MLIPs. During the inference stage, the Monte Carlo dropout method needs four independent runs to obtain uncertainty. GMM obtains uncertainty through iterative calculations using the expectation-maximization algorithm, and it also requires a longer time to compute uncertainty. Both MVE and eIP have minimal training and inference times, comparable to that of a normal MLIP. Regarding the force prediction accuracy shown in \cref{fig:silica}(c), ensemble, GMM, and eIP achieve the lowest errors, while dropout and MVE exhibit higher errors. \cref{fig:silica}(d) and (e) further illustrate the comparison of Spearman's correlation and ROC-AUC, respectively. 
Notably, \cref{fig:silica}(e) shows that eIP performs even better than the ensemble method on the ROC-AUC metric.

\subsection{Applications}\label{}

\begin{figure}[ht]
    \centering
    \includegraphics[width=12cm]{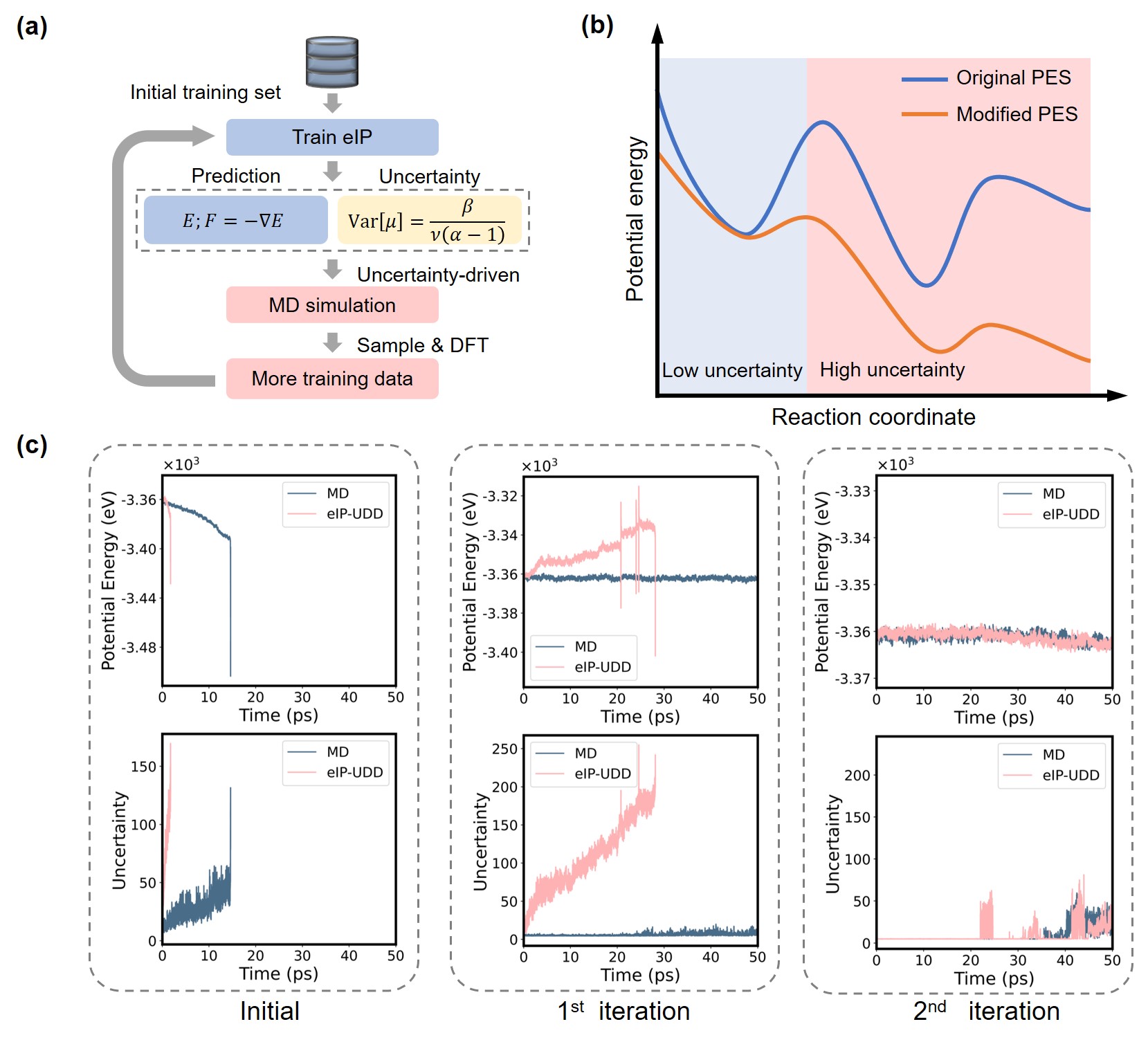}
    \caption{\textbf{Active learning with eIP.} (a) Workflow. Potential energy and uncertainty are calculated simultaneously by eIP. (b) Illustration of uncertainty-driven dynamics (UDD). The potential energy surface (PES) is adaptively modified according to uncertainty, with the potential energy in high-uncertainty regions being reduced to facilitate enhanced sampling. (c) Simulation results in each generation. The evolution of potential energy and uncertainty over time is shown for both convential MD and eIP-UDD simulations. In MD simulations, the PES remains unmodified, whereas in eIP-UDD simulations, the PES is modified based on the uncertainty from eIP. }
    \label{fig:water}
\end{figure}

\textbf{Active learning with eIP.} 
UQ plays a key role in active learning for training set construction. The quality of the training set is particularly crucial for MLIP, as the accuracy of MLIPs can significantly decrease when encountering unseen atomic configurations, leading to the collapse of simulations~\cite{fu2022forces}. \cref{fig:water}(a) illustrates a typical active learning workflow for MLIPs, where the data points with high uncertainty are iteratively explored to enrich the training set. In addition, uncertainty-driven dynamics (UDD) simulation~\cite{kulichenko2023uncertainty} can be employed to enhance sampling efficiency. In UDD simulations, potential energy surface is modified so that the atomic configurations with higher uncertainties are assigned lower potential energies, and consequently, these structures become more accessible, as indicated in \cref{fig:water}(b). The implementation of UDD simulation with eIP is provided in Methods.

We demonstrate the active learning process with eIP, using a water dataset as an example. In each generation, we performed both standard MD simulation and eIP-UDD simulation, and the changes in uncertainty and energy over simulation time are illustrated in \cref{fig:water}(c). 
The initial training set comprises 1,000 configurations sampled from a classical MD simulation trajectory generated using an empirical force field. The abnormal energy fluctuations suggest that both the MD and eIP-UDD simulations collapse very early. 
In the first iteration, the MD simulation remains stable after 50 ps. Although the eIP-UDD simulation collapses after 20 ps, the uncertainty increases over time, indicating that more previously unseen configurations are explored during the eIP-UDD simulation. 
In the second iteration, both the MD and the eIP-UDD simulations achieve stability after 50 ps. We also observe that the uncertainty does not increase significantly and this may suggest that configurations are explored sufficiently around certain local minima. 

\begin{figure}[ht]
    \centering
    \includegraphics[width=13cm]{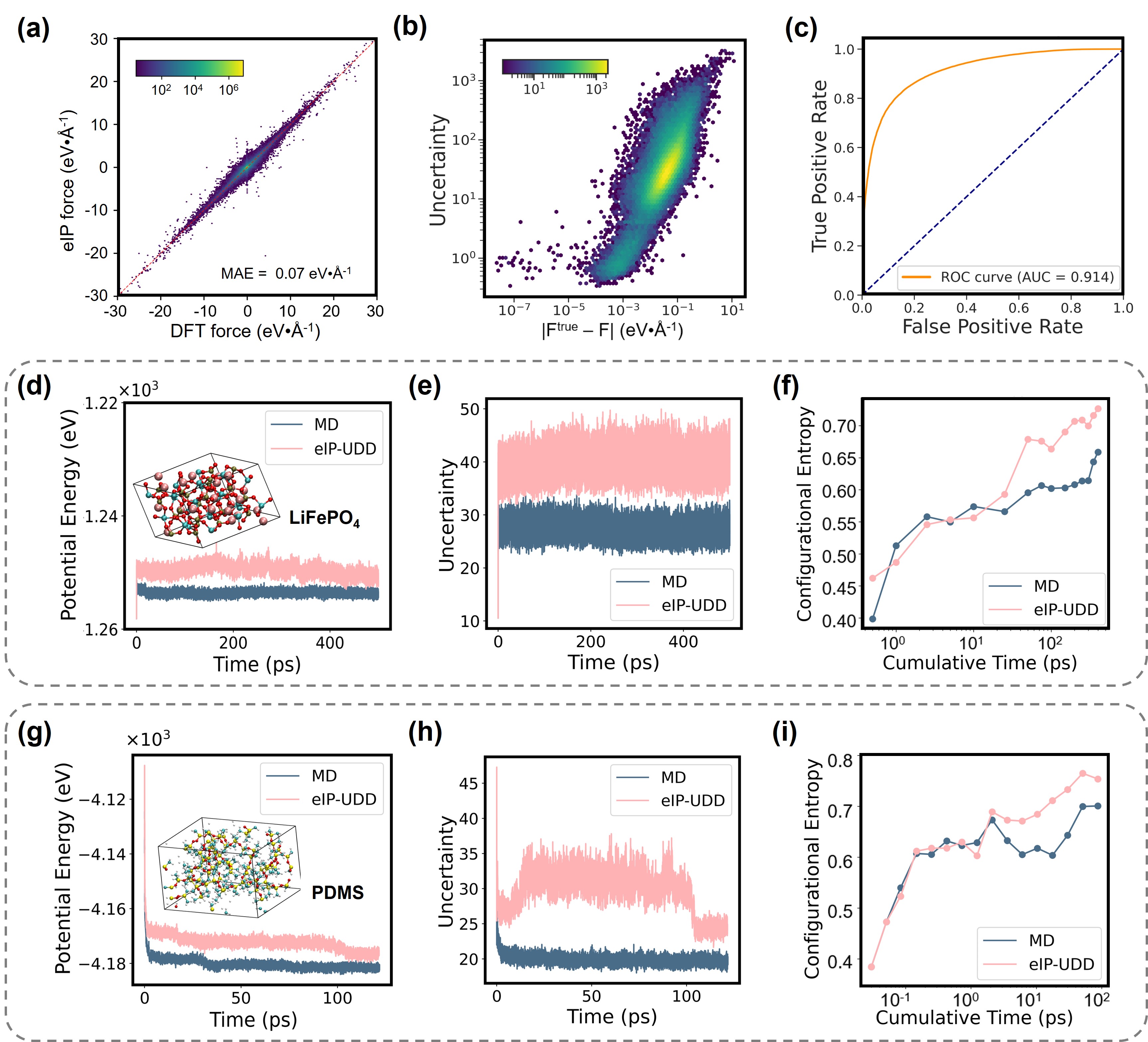}
    \caption{\textbf{Universal potential with eIP.} (a) Comparison of atomic forces between eIP prediction and ground truth. (b) Hexbin plots of uncertainties versus atomic force errors. The Spearman's rank correlation coefficient is 0.76. (c) ROC curve. The ROC-AUC score is 0.914.(d)-(f) Simulation results of $\text{LiFePO}_{4}$. 
    (g)-(i) Simulation results of polydimethylsiloxane (PDMS).
    The potential energy curves (d) and (g) indicate that both MD and eIP-UDD simulations are stable, demonstrating the effectiveness of the universal potential. The uncertainty curves (e) and (h) reveal that eIP-UDD configurations exhibit higher uncertainty levels for both materials. The evolutions of configurational entropy (f) and (i) further confirms that eIP-UDD simulations generate more diverse configurations than conventional MD simulations. 
    }
    \label{fig:universal}
\end{figure}

\textbf{Application of eIP in universal MLIP.} 
Finally, we explored the performance of eIP in universal MLIPs. To this end, we trained the model on the Materials Project Trajectory (MPtrj) dataset~\cite{deng2023chgnet}. The hexibin plots and the ROC curve in \cref{fig:universal} (a)-(c) demonstrate the performance of eIP on such a large dataset. 
Then we tested the performance of eIP in enhanced sampling using UDD simulation. 
We selected two distinct materials as examples, namely lithium iron phosphate ($\text{LiFePO}_{4}$) and polydimethylsiloxane (PDMS). 
$\text{LiFePO}_{4}$ is a mature commercial cathode material for lithium ion batteries, while PDMS is a widely applied organosilicon polymer material. These materials serve as benchmarks for evaluating the configurational sampling performance of eIP-UDD simulations for both inorganic crystalline and organic polymeric systems. 
For each material, changes in potential energy, uncertainty, and configurational entropy over simulation time are shown in \cref{fig:universal}(d)-(i). 
As expected, the trajectory of the eIP-UDD simulation has a larger uncertainty than that of the conventional MD simulation. 
The results of the configurational entropy in \cref{fig:universal}(f) and (i) further prove that the eIP-UDD simulations have obtained more diverse configurations.

\subsection{Discussions}\label{}
UQ is a critical topic in various fields of machine learning, particularly in scientific applications such as molecular simulations based on MLIP. Conventional UQ methods suffer from either high computational costs or decreased prediction accuracy. In this work, we propose a single-model UQ method, called eIP, which achieves both efficiency and accuracy, as demonstrated by extensive experiments in various applications. The eIP framework incorporates locality, directionality, and quantile regression, all of which are essential for achieving optimal results. This is evident from the ablation study presented in Supplementary S3, where the absence of any single component leads to a noticeable decline in performance. 

Although ensemble methods have been widely used in active learning, they typically require training four or more models simultaneously. In practice, this process usually involves dozens or more iterations and takes a significant amount of time and computational resources to obtain a satisfactory training set. As a result, single-model UQ methods, such as eIP, have the potential to save several months in applications, making eIP a more efficient alternative when time constraints and computational resources are a significant concern. In addition, for large-scale simulations, ensemble methods require a significant amount of computation to evaluate the reliability of MLIP-based MD simulations, while eIP facilitates real-time assessment without incurring noticeable additional costs. 

\section{Methods}\label{sec3}

\subsection{Formulism of eIP}\label{}
We employ quantile regression with maximum likelihood estimation to better model the uncertainty of MLIPs. Quantile regression is solved by minimizing the tiled loss for a given quantile $q$:

\begin{equation}
\mathcal{L}_i = \rho_q(\epsilon_i) = \max(q\epsilon_i,(q-1)\epsilon_i),
\label{eq:tiled}
\end{equation}

where $\epsilon_i$ denotes the residue for observation $i$. 

The quantile $q$ follows an asymmetric Laplace distribution with mean $\mu$, variance $\sigma$, and an asymmetrical parameter equal to the quantile $q$~\cite{yu2005three}. The likelihood function can be expressed as a scalar mixture of Gaussians~\cite{kotz2012laplace, kozumi2011gibbs} $\mathcal{N}(\mu+{\tau}z, \omega\sigma z)$, where $\tau = \frac{1-2q}{q(1-q)}$, $\omega = \frac{2}{q(1-q)}$, $z \sim \text{exp}\left(\frac{1}{\sigma}\right)$. 

We assume that the atomic forces $F \in \mathbb{R}^{N\times3}$ come from a Gaussian distribution, but the mean and variance are unknown. For instance, the x-component of the force on the atom $i$ follows:

\begin{equation}
f_{ix} \sim \mathcal{N}(\mu_{ix} + \tau z_{ix}, \omega \sigma_{ix} z_{ix}).
\end{equation}

By placing a Gaussian prior on the unknown mean $\mu_{ix}$ and an Inverse-Gamma prior on the unknown variance $\sigma_{ix}$, we obtain the Normal-Inverse-Gamma evidential prior $p(\mu_{ix},\sigma_{ix}|\mathbf{m}_{ix})$ with a set of parameters $\mathbf{m}_{ix}=(\gamma_{ix},\nu_{ix},\alpha_{ix},\beta_{ix})$~\cite{amini2020deep, huttel2023deep}. 
As a result, $\gamma$ is equal to the predicted force

\begin{equation}
    \mathbb{E}[\mu_{ix}]=\gamma_{ix},
\end{equation}

and the x-component of epistemic uncertainty for the atom $i$ is

\begin{equation}
\text{Var}[\mu_{ix}] = \frac{\beta_{ix}}{\nu_{ix} (\alpha_{ix}-1)}.
\end{equation}

The y- and z-components are computed similarly. We define the uncertainty $\sigma_i$ associated with the atom $i$ as

\begin{equation}
\sigma_i^2 = \sqrt{
\left(\frac{\beta_{ix}}{\nu_{ix} (\alpha_{ix}-1)}\right)^2 +
\left(\frac{\beta_{iy}}{\nu_{iy} (\alpha_{iy}-1)}\right)^2 +
\left(\frac{\beta_{iz}}{\nu_{iz} (\alpha_{iz}-1)}\right)^2
}.
\end{equation}

The uncertainty for a configuration composed of $N$ atoms is determined by computing the average:

\begin{equation}
    \sigma = \frac{1}{N} \sum_{i=1}^N \sigma_i.
\end{equation}

The parameter $\gamma_{ix}$ is equal to the predicted force $f_{ix}$, which is computed as the negative gradient of the predicted potential energy $E$. Other parameters, $\nu_{ix}$, $\alpha_{ix}$, and $\beta_{ix}$, are inferred by neural networks based on their corresponding atomic features. The model is trained by maximizing the probability $p(f_{ix}|\mathbf{m}_{ix})$, leading to the negative log-likelihood (NLL) loss function~\cite{huttel2023deep}:

\begin{equation}
\begin{aligned}
\mathcal{L}_{ix}^\text{NLL} = & \frac{1}{2} \log\left(\frac{\pi}{\nu_{ix}}\right) - \alpha_{ix} \log(\Omega) \\
& + \left(\alpha_{ix} + \frac{1}{2}\right) 
\log\left((f_{ix}^\text{true} - (\gamma_{ix} + \tau z_{ix}))^2 \nu_{ix} + \Omega\right) \\
& + \log\left(\frac{\Gamma(\alpha_{ix})}{\Gamma(\alpha_{ix} + \frac{1}{2})}\right).
\end{aligned}
\label{eq:nll}
\end{equation}

where $\Omega=4\beta_{ix}(1+\omega z_{ix}\nu_{ix})$, $z_{ix}=\frac{\beta_{ix}}{\alpha_{ix}-1}$, and $\Gamma(\cdot)$ is the gamma function. 

We use an evidence regularizer so that the model tends to output low confidence when the predictions are incorrect:

\begin{equation}
\mathcal{L}_{ix}^\text{R} = \rho_q(f_{ix}^\text{true}-\gamma_{ix}) \cdot \left(2\nu_{ix} + \alpha_{ix} + \frac{1}{\beta_{ix}}\right).
\end{equation}

The y- and z-components are computed similarly. Finally, the overall loss function, including the L1 loss for energy prediction, is:

\begin{equation}
\mathcal{L}  = |E^\text{true}-E| + \frac{w}{3N}\sum_{i=1}^N\sum_{a\in(x,y,z)} \left(
{\mathcal{L}_{ia}^\text{NLL} + \lambda \mathcal{L}_{ia}^\text{R}} \right),
\label{loss}
\end{equation}

where $w$ and $\lambda$ are hyperparameters to adjust the weighting of each term. 
The details of eIP implementations are provided in Supplementary S4.

\subsection{Datasets}\label{}

\textbf{ISO17 dataset.}
The ISO17 dataset~\cite{schnet} was obtained from \url{http://quantum-machine.org/datasets/}. We adopted the original splitting strategy for the training, validation, and test set. For training sets of different sizes, the smaller training sets were randomly sampled from the largest training set containing 400,000 conformations. 

\textbf{Silica glass dataset.}
The silica glass dataset is obtained from a previously published study~\cite{tan2023single}. The dataset comprises 1691 configurations, each containing 699 atoms (233 Si and 466 O atoms), and we adopted the original dataset splitting scheme for training, validation, and testing. These configurations are generated through molecular dynamics simulations under various conditions, and density functional theory (DFT) calculations are performed to obtain the energies and forces. 

\textbf{Water dataset.}
The initial water training set is taken from our previous work~\cite{cui2024online}. It comprises 1,000 configurations sampled from classical MD trajectories with an empirical force field. Each configuration contains 288 atoms with periodic boundary conditions. 
During active learning, we ran UDD simulations at 300 K and sampled 1,000 configurations for each iteration. The energies and forces are determined using density functional theory (DFT) calculations employing the cp2k software package~\cite{kuhne2020cp2k} with the PBE-PAW-DFT-D3 method~\cite{perdew1996generalized, blochl1994projector, grimme2010consistent}. 

\textbf{MPtrj dataset.}
The MPtrj dataset~\cite{deng2023chgnet} is a collection of MD trajectories designed for training a universal potential. It comprises millions of configurations covering 89 elements and the energies and forces are determined using DFT calculations. We adopted the original splitting strategy with an 8:1:1 training, validation, and test ratio.

\subsection{Evaluation metrics}

\textbf{Spearman's rank correlation coefficient.}
Spearman's rank correlation is a non-parametric measure of the strength and direction of association between two ranked variables. Unlike Pearson's correlation, which accesses linear relationships, Spearman's rank correlation evaluates how well the relationship between two variables can be described using a monotonic function. We expect a larger error to be associated with a larger uncertainty, and their correlation does not necessarily be linear. Therefore, the Spearman's rank correlation coefficient was used to assess the reliability of the uncertainty. A coefficient of 1 means perfect correlation, and a coefficient of 0 indicates that there is no correlation between the ranks of the two variables. 

\textbf{Area under the receiver operating characteristic curve.}
The receiver operating characteristic (ROC) curve is a graphical representation of a classifier's performance. 
The area under the ROC curve (ROC-AUC) provides a complementary evaluation metric for UQ that avoids the possible limitations of using the Spearman's rank correlation coefficient alone. 
Following the approach of a previous study~\cite{tan2023single}, we designed a classification task in which predictions with high errors are expected to exhibit high levels of uncertainty. 
The ROC-AUC score ranges from 0 to 1, with a score of 1 denoting a perfect classifier and 0.5 indicating performance no better than random choice.

\textbf{Configurational entropy.}
Configurational entropy quantifies the number of ways that atoms in a system can be arranged. High entropy indicates that the system is likely to take on many different arrangements, whereas low entropy implies a more ordered, less random state. We used configurational entropy as a metric to measure the diversity of configurations obtained during MD and UDD simulations. The formula for configurational entropy is:

\begin{equation}
    S_\text{conf} = -\sum_t p(\mathcal{C}_t)\log(p(\mathcal{C}_t)),
\end{equation}

where $p(\mathcal{C}_t)$ is the probability distribution of a configuration at timestep $t$. 
We estimated the probability distribution using the histogram of order parameters. 
For LiFePO$_4$, the selected order parameters were the P-O-Fe angle and the PO$_4$ tetrahedral distortion. For PDMS, we selected the end-to-end distance and the radius of gyration as the order parameters. 
To determine the probability distribution, the order parameter space was discretized into an $N_\text{e} \times N_\text{e}$ grid, and the frequency of configurations within each grid cell was calculated. The configurational entropy was normalized by dividing it by the maximum possible entropy value, $2\log(N_\text{e})$, resulting in values between 0 and 1. A larger grid size $N_\text{e}$ offers a finer resolution but may suffer from statistical noise, while a smaller $N_\text{e}$ provides more robust statistics at a lower resolution. We used $N_\text{e}=40$ for all reported results. Varying the value of $N_\text{e}$ does not significantly affect the results, as the configurational space was sampled sufficiently in our simulations.

\subsection{Molecular dynamics (MD) simulations}\label{}

MD simulations were performed using the Atomic Simulations Environment (ASE) Python library~\cite{larsen2017atomic}. 
The simulations are set with a timestep of 0.1 fs in the canonical (NVT) ensemble. The Berendsen thermostat~\cite{berendsen1984molecular} was used with a coupling temperature of 300 K and a decaying time constant $\tau$ of 100 fs. The atomic velocities were initialized according to the Boltzmann distribution at 300K. 
The initial water configuration was selected from the water test set. The LiFePO$_4$ configuration was obtained from the Materials Project, comprising 168 atoms in the unit cell. The PDMS configuration was constructed using three polymer chains with a polymerization degree of 25 and a density of 0.97 $\text{g} \cdot \text{cm}^{-3}$, containing 759 atoms in total. All systems were modeled with periodic boundary conditions.

\subsection{Uncertainty-driven dynamics (UDD) simulations}\label{}
The UDD simulation technique utilizes a bias energy that favors configurations with higher uncertainties. Kulichenko et al. introduce a bias energy~\cite{kulichenko2023uncertainty} defined as:
\begin{equation}
E_\text{bias}(\sigma^2) = A\left[\exp\left(-\frac{\sigma^2}{NB^2} \right)-1\right],
\end{equation}
where the parameters $A$ and $B$ are chosen empirically. 
The bias force $F_\text{bias}$ is then determined by calculating the negative gradient of the bias energy:
\begin{equation}
F_\text{bias} = -\nabla(E_\text{bias}(\sigma^2))
= -E_\text{bias}(\sigma^2)' \nabla \sigma^2.
\label{eq:E_bias}
\end{equation}
By leveraging eIP for UQ, the gradient of $\sigma$ can be obtained through automatic differentiation. 

Notably, the bias force could become exceptionally large, leading to the collapse of molecular simulations. 
We found that limiting the magnitude of the bias forces using a clipping strategy proved not effective. 
To prevent this issue, we incorporate a Gaussian term to limit the magnitude of the bias force with two additional empirically chosen parameters $C$ and $D$: 
\begin{equation}
  F_\text{bias}^\text{limited}= F_\text{bias}\frac{D}{\sqrt{2\pi}C}\exp\left(\frac{-F_\text{bias}^2}{2C^2}\right).
\label{eq:F_bias}
\end{equation}
This adjustment of bias force implies a new bias energy formulation and ensures more stable UDD simulations. 
Detailed discussions about the empirical parameters $A$, $B$, $C$, and $D$ are provided in the Supplementary Section S6. 
Finally, the combined force $F + F_\text{bias}^\text{limited}$ is used to guide the simulations toward configurations with higher uncertainties, enhancing the sampling for more diverse atomic configurations.

\section*{Data availability}
The training data used for all models in this work are publicly available. The generated checkpoints and simulation trajectories are available at figshare~\cite{xu2025}.

\section*{Code availability}
The source code for reproducing the key findings in this work is available at \url{https://github.com/xuhan323/eIP}. 

\section*{Acknowledgments}
This work was supported by Shanghai Artificial Intelligence Laboratory, Shanghai Committee of Science and Technology, China (Grant No. 23QD1400900), and the National Natural Science Foundation of China (Grant No. 12404291). H.X., T.C., and T.C. did this work during their internship at Shanghai Artificial Intelligence Laboratory. 

\section*{Author contributions}
M.S. and S.Z. conceived the idea and led the research. H.X. and T.C. developed the eIP code and trained the models. H.X. and J.M. performed the experiments and analyses. C.T. developed the active learning workflow and performed the molecular dynamics simulations. Y.L., X.G., and X.G. contributed technical ideas for datasets and experiments. D.Z. and W.O. contributed technical ideas for designing and training the models. H.X, C.T., and M.S. wrote the first draft. All authors discussed the results and reviewed the manuscript. 

\section*{Competing interests}
The authors declare no competing interests.


\bibliography{sn-bibliography}

\end{document}




\thispagestyle{empty}
\tableofcontents
\listoftables
\listoffigures

\clearpage

\pagenumbering{arabic}

\section{A toy system}
\begin{figure}[ht]
    \centering
    \includegraphics[width=13cm]{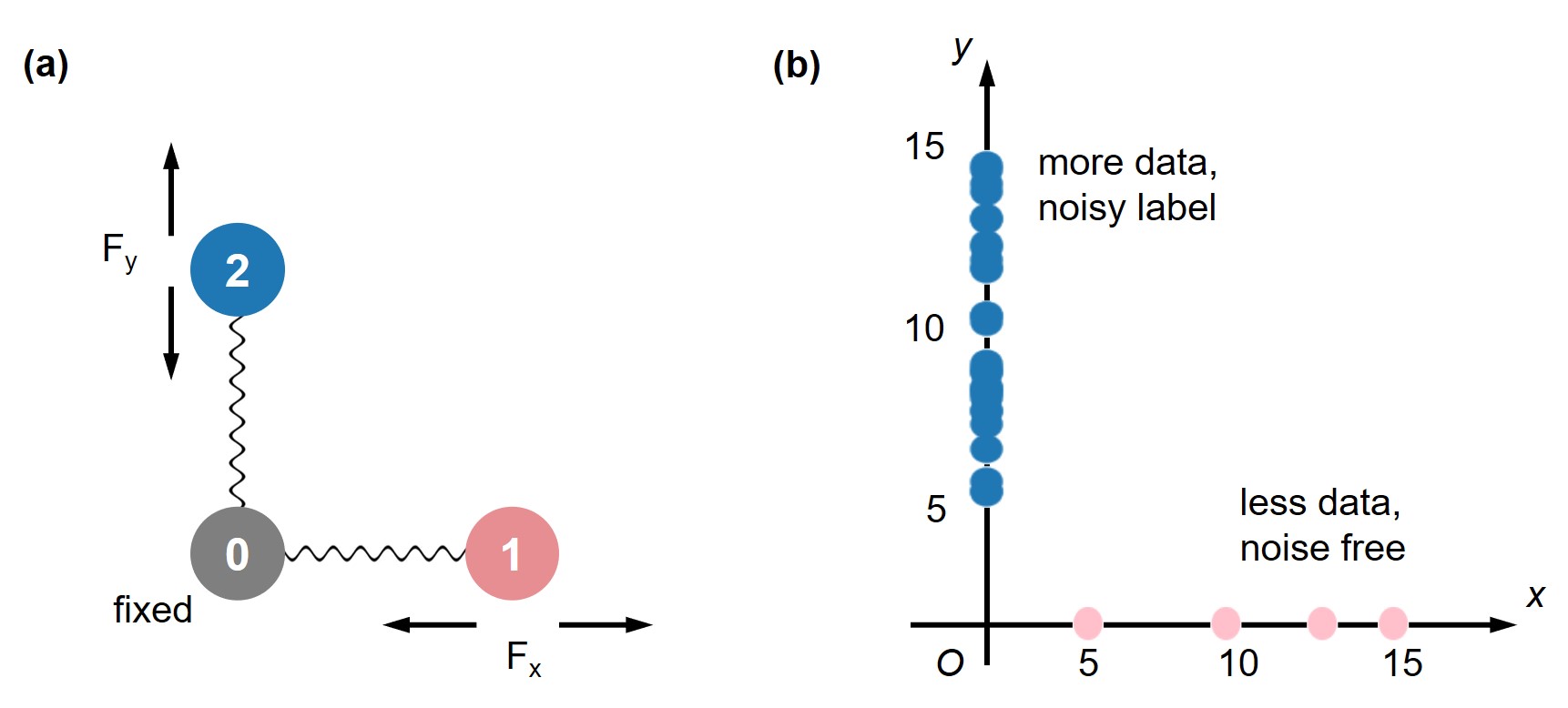}
    \caption[The three-atom system]{\textbf{The three-atom system.} (a) Atom 0 is fixed at the origin, atom 1 can only move in the x direction, and atom 2 can only move in the y direction. (b) Illustration of the distribution of atoms 1 and 2 in the training set. }
    \label{fig:toy}
\end{figure}

We introduce a toy system to demonstrate the locality and directionality of uncertainty in atomic systems. As shown in \cref{fig:toy}(a), consider a system composed of three atoms: atom 0 is fixed at the origin, atom 1 can only move in the x direction, and atom 2 can only move in the y direction. Ignoring the interaction between atoms 1 and 2 and assuming bonded interactions between the other atom pairs, the x-component of the force on atom 0 ($F_x$) depends solely on atom 1, and the y-component of the force on atom 0 ($F_y$) depends solely on atom 2: 
\begin{align}
    F_x=k*(x-x_0),\\
    F_y=k*(y-y_0),
\end{align}
where we set the parameters $k=5$, $x_0=10$, and $y_0=10$. The $x$ and $y$ denote the x-coordinate of atom 1 and the y-coordinate of atom
2 respectively. The specific values are chosen for illustration purposes and all units are ignored. 

In this three-atom system, only the x-coordinate of atom 1 and the y-coordinate of atom 2 are variable, with their values randomly selected between 5 and 15. The forces were computed using the formulas above. To investigate eIP' ability to disentangle epistemic and aleatoric uncertainties, there are restrictions when generating the coordinates for the training data, as shown in \cref{fig:toy}(b). Firstly, the x-coordinate (atom 1) are limited to only four distinct values (rather than being evenly distributed between 5 and 15), while allowing the y-coordinate (atom 2) to vary freely between 5 and 15. Secondly, we introduced Gaussian random noise to the y-component of the force (atom 2). As a result, the epistemic uncertainty for $F_x$ is expected to be large, while that for $F_y$ should be small. On the contrary, the aleatoric uncertainty for $F_x$ should be small, while that for $F_y$ should be large. 

\begin{figure}[ht]
    \centering
    \includegraphics[width=13cm]{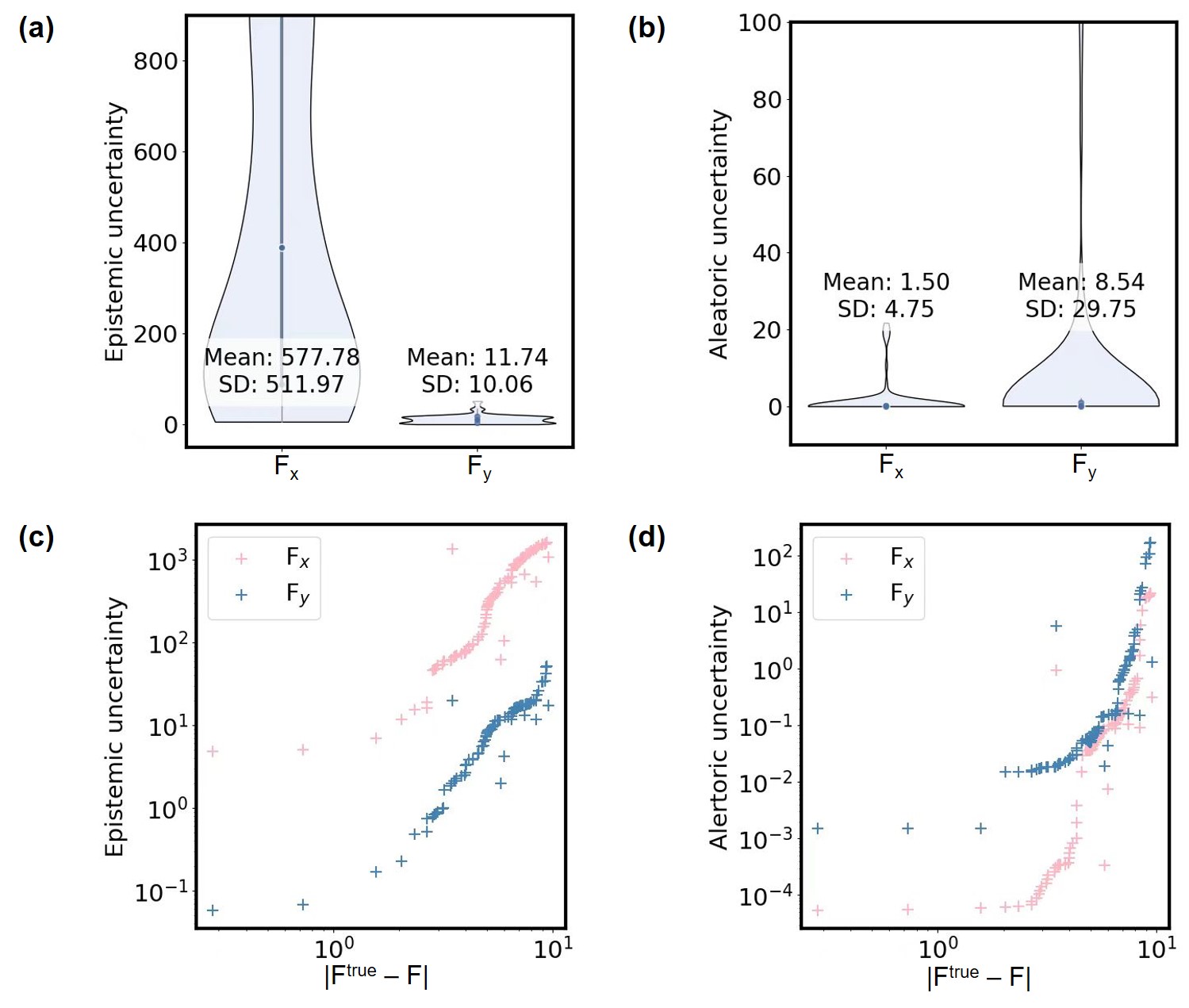}
    \caption[Results on the three-atom system]{\textbf{Results on the three-atom system.} (a) Violin plot of epistemic uncertainty. (b) Violin plot of aleatoric uncertainty. (c) Scatter plot of epistemic uncertainty versus force error. (d) Scatter plot of aleatoric uncertainty versus force error.}
    \label{fig:toy_system}
\end{figure}

We generated 5,000 configurations, divided into training, validation, and test sets with a 6:2:2 ratio. The results are shown in \cref{fig:toy_system}, where the uncertainty varies in different directions, and the distributions of aleatoric and epistemic uncertainties align with our expectations. 

\clearpage

\section{Aleatoric uncertainty}
An advantage of eIP is its ability to distinguish between aleatoric and epistemic uncertainty. Aleatoric uncertainty stems from noise in the dataset, and larger noise leads to larger prediction errors. While both aleatoric and epistemic uncertainty show a positive correlation with model prediction errors, we can effectively decouple these two types of uncertainty through experimental designs. This can be achieved by deliberately introducing small amounts of noise to the labels while maintaining consistent model error rates. A similar approach was adopted by Caldeira et al. in their study of uncertainty in pendulum motion problems~\cite{caldeira2020deeply}.

To verify eIP's performance in estimating aleatoric uncertainty, we introduce Gaussian noise to the training data labels. As an example, we use the aspirin molecules from the MD17 dataset~\cite{chmiela2017machine}. We selected 1,000 aspirin conformations for each of the following: the training set, the validation set, and the test set. 
We introduce Gaussian noise with standard deviations of 0.01 kcal mol$^{-1}$\AA$^{-1}$ and 0.1 kcal mol$^{-1}$\AA$^{-1}$, respectively, to the forces of the training data. We train the models on datasets with different noise levels and validate/test them on the same noiseless validation/test dataset. \cref{fig:aleatoric}(a) shows the scatter plot of aleatoric uncertainties versus force errors, indicating that the aleatoric uncertainty increases with the noise level. 
In contrast, the epistemic uncertainty does not change much with the increase in noise level, as shown in \cref{fig:aleatoric}(b). We also trained on the dataset with a larger noise level (1.0 kcal mol$^{-1}$\AA$^{-1}$) and found that the prediction error and both types of uncertainties increased significantly.

\begin{figure}[ht]
    \centering
    \includegraphics[width=13cm]{figure/al&ep.jpg}
    \caption[Aleatoric and epistemic uncertainties]{\textbf{Aleatoric (a) and Epistemic (b) uncertainties.} Gaussian noise with different standard variations $\sigma$ is introduced to the forces of training data. The aleatoric uncertainty increases with the noise level, while the epistemic uncertainty does not change much with the increase in noise level.}
    \label{fig:aleatoric}
\end{figure}

\clearpage

\section{Ablation Study}
Using the aspirin dataset, we also conducted ablation experiments to investigate the significance of locality, directionality, and quantile regression in the design of eIP. When ablating locality, we chose global embedding to estimate the uncertainty associated with the energy. When ablating directionality, we replace the equivariant backbone PaiNN with the invariant backbone SchNet~\cite{schnet}. When ablating quantile regression, we utilize standard deep evidential regression. 

\begin{figure}[ht]
    \centering
    \includegraphics[width=13cm]{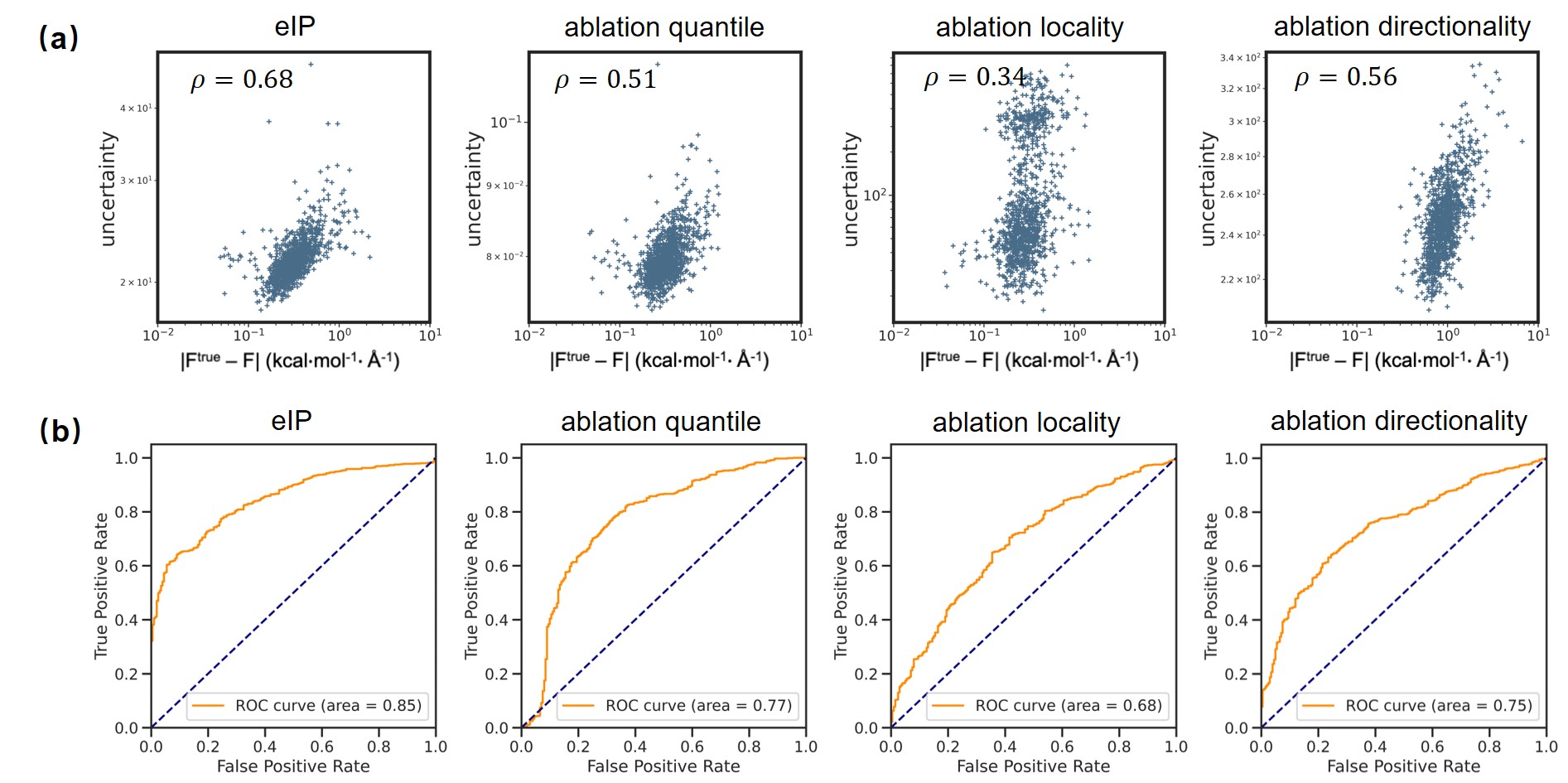}
    \caption[Ablation experiment]{\textbf{Ablation experiment.} (a) Scatter plot of epistemic uncertainty versus force error. (b) ROC curves. The Spearman's rank correlation coefficients $\rho$ and ROC-AUC scores are shown in the insets. }
    \label{fig:ablation1}
\end{figure}

The scatter plots of epistemic uncertainties versus force errors in \cref{fig:ablation1}(a) reveal the effect of each component's ablation. 
The ROC curves in \cref{fig:ablation1}(b) further demonstrate that the ablation of each component leads to measurable decreases in the performance of eIP.

\clearpage

\section{Method details}
The eIP model in this work is implemented using the PaiNN architecture as the backbone. For the MPtrj dataset, we employed four message-passing layers with a hidden dimension of $d=256$. For the other datasets, we used three message-passing layers with a hidden dimension of $d=128$. 

In contrast to the standard PaiNN model, the eIP model incorporates additional fully connected layers. These layers take the equivariant features as input and produce the output $\alpha$, $\beta$, and $\nu$. 
Specifically, for a system composed of $N$ atoms, starting with the extracted equivariant feature $V\in\mathbb{R}^{N\times3\times d}$, we compute: 

\begin{equation}
    T=W_2\left( \text{shifted\_softplus}(W_1V+b_1) + b_2\right),
\end{equation}

where $W_1\in\mathbb{R}^{d\times d}$, $W_2\in\mathbb{R}^{d\times 3}$, $b_1\in\mathbb{R}^d$, and $b_2\in\mathbb{R}^3$ are trainable parameters. The resulting tensor $T\in\mathbb{R}^{N\times3\times3}$ is then split along its last dimension:

\begin{equation}
    T=\left[T^\alpha, T^\beta, T^\nu\right],
\end{equation}

where $T^\alpha\in\mathbb{R}^{N\times3}$, $T^\beta\in\mathbb{R}^{N\times3}$, and $T^\nu\in\mathbb{R}^{N\times3}$. To constrain the range of values and ensure numerical stability, we apply the softplus function and introduce a small constant to obtain the output:
\begin{align}
    \alpha &= \text{softplus}(T^\alpha)+1+\varepsilon,\\
    \beta &=  \text{softplus}(T^\beta),\\
    \nu &= \text{softplus}(T^\nu)+\varepsilon.
\end{align}


In the eIP loss function, we introduced three additional hyperparameters: $\lambda$, $q$, and $w$. The hyperparameter $\lambda$ balances the trade-off between uncertainty inflation and model fit. Setting $\lambda$ too small results in an overconfident estimate, while setting it too high leads to over-inflation. Empirically, setting $\lambda$ to 0.1 yields good results across all of our experiments. The hyperparameter $w$ adjusts the relative importance between energy and force. We observed that a small $w$ may result in training failure; therefore, using a larger value for $w$ is recommended. The hyperparameters used for each dataset are listed in \cref{tb:hyper}. 

\vspace{2em}

\begin{table*}[h]
\centering
\renewcommand\arraystretch{1.2}
\setlength{\abovecaptionskip}{0pt}
\setlength{\belowcaptionskip}{10pt} 
\caption[Hyperparameters]{Hyperparameters.}
\label{tb:hyper} 
\begin{tabular}{ccccc}
\bottomrule   & ISO17 & silica glass & water & MPtrj \\
\hline
cutoff  & 5.0 & 5.0 & 5.0 & 6.0\\
batch size   & 16 & 2 & 2 & 2\\
initial lr  & 5e-4 & 1e-4 & 5e-4 & 1e-4\\
lr strategy & StepLR& ReduceOnPlateau& StepLR& ReduceOnPlateau\\
lr decay ratio  & 0.5 & 0.5 & 0.5 & 0.5\\
lr step size & 150 & 30 & 150 & 30\\
max epochs  & 1000 & 2000 & 1000 & 1000\\
$\lambda$  & 0.1 & 0.1 & 0.1 & 0.1\\
$w$  & 1e5 & 1e5 & 10 & 10\\
$q$  & 0.6 & 0.6 & 0.6 & 0.4\\
\bottomrule
\end{tabular}
\end{table*}


\clearpage

\section{Other uncertainty quantification methods}
\textbf{Ensemble.}
The ensemble method~\cite{breiman1996bagging} achieves uncertainty quantification by combining predictions from multiple models. Those models have the same architecture but are initialized differently during training. Therefore, the consistency of model predictions can be used as a measure of confidence. Taking into account the predictions of the atomic force, the uncertainty $\sigma_{ix}$ of the atom $i$ along the x-coordinate is computed as:

\begin{equation}
\sigma_{ix}^2 = \frac{1}{M}\sum_{m = 1}^{M} \left(F_{ix,m}-\overline{F_{ix}} \right)^2
\end{equation}

where $M$ is the number of models, $F_{ix,m}$ is the force predicted by the $m$-th model, and $\overline{F_{ix}}$ is the ensemble average. 
Since the uncertainty is a three-dimensional vector like the atomic force, we compute its norm to represent the uncertainty associated with an atom. 
\begin{equation}
\sigma_i = \sqrt{\sigma_{ix}^2 + \sigma_{iy}^2 + \sigma_{iz}^2}.
\end{equation}
The uncertainty of a configuration composed of $N$ atoms is obtained by calculating the average uncertainty over all atoms:
\begin{equation}
\sigma = \frac{1}{N}\sum_{i = 1}^{N} \sigma_i.
\end{equation}

\textbf{Monte Carlo Dropout.}
Dropout~\cite{srivastava2014dropout} is a technique used in neural networks to prevent over-fitting by randomly setting neurons to zero during training. Dropout is generally disabled during testing. However, in Monte Carlo dropout~\cite{gal2016dropout} for uncertainty quantification, dropout is enabled during both training and testing. 
By making multiple predictions with dropout, we will have a range of outputs instead of a single-value prediction. The variation in these outputs provides a measure for uncertainty quantification. 
Considering the atomic force predictions, the uncertainty $\sigma_{ix}$ of atom $i$ along the x-coordinate is computed as:

\begin{equation}
\sigma_{ix}^2 = \frac{1}{M}\sum_{m = 1}^{M} \left(F_{ix,m}-\overline{F_{ix}} \right)^2
\end{equation}

where $M$ is the number of times predictions are made, $F_{ix,m}$ is the force obtained from the $m$-th prediction, and $\overline{F_{ix}}$ is the average of the outputs. 
The uncertainties associated with an atom and a configuration are computed similarly as in the ensemble method. 

\textbf{Gaussian mixture model (GMM).}
In the GMM method, the distribution of atomic features is modeled as a weighted sum of $M$ Gaussians:

\begin{equation}  
p(x|\theta) = \sum_{m=1}^{M} w_m \mathcal{N}(x| \mu_m, \Sigma_m),  
\end{equation}  

where $x$ is the scalar feature extracted from the final layer of a trained MLIP, $w_m$ is the weight of the $m^\text{th}$ Gaussian, $\mu_m$ is the mean vector, and $\Sigma_m$ is the covariance matrix. The parameters $\theta=\{w_m,\mu_m,\Sigma_m\}$ are optimized using the expectation-maximization (EM) algorithm, and the number of Gaussians $M$ is determined using the Bayesian Information Criterion (BIC). To obtain uncertainty, we extract the feature vector of a test atom $\hat{x}$ and compute the negative log-likelihood (NLL):

\begin{equation}  
\text{NLL}(\hat{x}|X) = -\log \left( \sum_{m=1}^{M} w_m \mathcal{N}(x| \mu_m, \Sigma_m) \right),  
\end{equation}  

where $X$ denotes the atomic features of all training data. The NLL value directly serves as an atomic uncertainty metric, with higher NLL values indicating higher uncertainties.  

\textbf{Mean-variance estimation (MVE).}
In the MVE method, the neural network predicts a distribution $\mathcal{N}(\mu,\sigma)$ instead of the target itself. This method employs maximum likelihood estimation to train the models. Here, MVE was applied only to the force prediction. Specifically, the model outputs the energy $E$ and the variance of force $\sigma^2_{F_{ia}}$, and the mean of force $\mu_{F_{ia}}$ is computed via the gradient of $E$. The loss function is:

\begin{equation}
\mathcal{L} = |E^\text{true}-E| + w\left(\frac{1}{2}\frac{1}{3N}\sum_{i = 1}^{N}\sum_{a\in(x,y,z)}\log(2\pi\sigma^2_{F_{ia}})+\frac{(F_{ia}-\mu_{F_{ia}})^2}{2\sigma^2_{F_{ia}}}\right),
\end{equation}

where the hyperparameter $w$ regulates the trade-off between the energy and force predictions. The predicted variances $\sigma^2_{F_{ia}}$ serve as the atomic uncertainty metric.

\clearpage

\section{Uncertainty-driven dynamics (UDD)}

The use of UDD simulations enables more efficient sampling of configuration space. By leveraging the epistemic uncertainty quantified by eIP, we modified the potential energy surface and performed UDD simulations. 

Our implementation of eIP involves four adjustable parameters in the bias force: $A$, $B$, $C$, and $D$ (see Methods in the main text). Selecting appropriate parameters is crucial but manageable for UDD simulations. As a practical guideline, $A$ is chosen to be of the same magnitude as the potential energy $E$. This ensures that the bias energy $E_\text{bias}(\sigma^2)$ remains proportional and does not overwhelm the system dynamics. The value of $B$ controls the exponential decay rate of the bias energy. A larger $B$ slows the decay, leading to a broader response to changes in $\sigma$. Its value should be balanced to capture rapid fluctuations without compromising the stability of the bias energy.
For structurally stable molecules that require larger bias potentials to escape local minima, $A$ should be increased while keeping $NB^2$ (denominator) and $\sigma^2$ (numerator) numerically close.
Parameters $C$ and $D$ regulate the decay and magnitude of the limited bias force $F_\text{bias}^\text{limited}$. A larger $C$ results in a faster decay, limiting $F_\text{bias}$ at larger values, while $D$ scales the overall magnitude of $F_\text{bias}^\text{limited}$. 

The specific values of $A$, $B$, $C$, and $D$ used in this work are listed in \cref{tb:udd}. 
The results of water UDD simulations with varying $A$, $B$, $C$, and $D$ are shown in \cref{fig:umap_a} to \cref{fig:umap_d}. 

\vspace{2em}

\begin{table*}[h]
\centering
\renewcommand\arraystretch{1.2}
\setlength{\tabcolsep}{8mm}
\setlength{\abovecaptionskip}{0pt}
\setlength{\belowcaptionskip}{10pt} 
\caption[UDD parameters]{UDD parameters.}
\label{tb:udd}
\begin{tabular}{c c c c c }
\bottomrule   & $A$ & $B$ & $C$ & $D$ \\
\hline
water  & 2000 & 20 & 1 & 7\\
LiFePO$_4$   & 5000 & 35 & 5 & 50\\
PDMS  & 10000 & 25 & 5 & 200\\

\bottomrule
\end{tabular}
\end{table*}

\begin{figure}[htb]
    \centering
    \includegraphics[width=13cm]{old/udd_a.png}
    \caption[Results of UDD with different parameter A]{\textbf{Results of UDD with different parameter A.} Different colored curves represent different A values when $B = 20$, $C = 1.0$, and $D = 7.0$. We use the model trained after four active learning iterations. (a) The variation of force average over simulation time. (b) The variation of force standard deviation over simulation time. (c) The variation of uncertainty average over simulation time. (d) The variation of uncertainty standard deviation over simulation time. }
    \label{fig:umap_a}
\end{figure}

\begin{figure}[htb]
    \centering
    \includegraphics[width=13cm]{old/udd_b.png}
    \caption[Results of UDD with different parameter B]{\textbf{Results of UDD with different parameter B.} Different colored curves represent different B values when $A = 10,000$, $C = 1.0$, and $D = 7.0$. We use the model trained after four active learning iterations. (a) The variation of force average over simulation time. (b) The variation of force standard deviation over simulation time. (c) The variation of uncertainty average over simulation time. (d) The variation of uncertainty standard deviation over simulation time.  }
    \label{fig:umap_b}
\end{figure}

\begin{figure}[htb]
    \centering
    \includegraphics[width=13cm]{old/udd_c.png}
    \caption[Results of UDD with different parameter C]{\textbf{Results of UDD with different parameter C.} Different colored curves represent different C values when $A = 10,000$, $B = 20.0$, and $D = 7.0$. We use the model trained after four active learning iterations. (a) The variation of force average over simulation time. (b) The variation of force standard deviation over simulation time. (c) The variation of uncertainty average over simulation time. (d) The variation of uncertainty standard deviation over simulation time.  }
    \label{fig:umap_c}
\end{figure}

\begin{figure}[htb]
    \centering
    \includegraphics[width=13cm]{old/udd_d.png}
    \caption[Results of UDD with different parameter D]{\textbf{Results of UDD with different parameter D} Different colored curves represent different D values when $A = 10,000$, $B = 20$, and $C = 1.0$. We use the model trained after four active learning iterations. (a) The variation of force average over simulation time. (b) The variation of force standard deviation over simulation time. (c) The variation of uncertainty average over simulation time. (d) The variation of uncertainty standard deviation over simulation time. }
    \label{fig:umap_d}
\end{figure}




\bibliography{sn-bibliography}